\documentclass[aps,prb,twocolumn,floatfix,showpacs]{revtex4}
\usepackage[dvips]{graphicx}
\usepackage{array}
\usepackage{bm,amssymb,amsmath}

\usepackage{epsf}

\begin{document}


\title{Unified phase diagram of models exhibiting neutral-ionic transition}

\author{\"O.~Legeza, K.~Buchta, and J.~S{\'o}lyom}

\affiliation{Research Institute for Solid State Physics and Optics, H-1525
Budapest, P.\ O.\ Box 49, Hungary }

\date{\today}

\begin{abstract}
We have studied the neutral-ionic transition in organic mixed-stack compounds.
A unified model has been derived which, in limiting cases, is equivalent to the 
models proposed earlier, the donor-acceptor model and the ionic Hubbard model. 
Detailed numerical calculations have been performed on this unified model with 
the help of the density-matrix renormalization-group (DMRG) procedure calculating 
excitation gaps, ionicity, lattice site entropy, two-site entropy, and the dimer order 
parameter on long chains and the unified phase diagram has been determined.
\end{abstract}


\maketitle

\section{Introduction}

Two physically seemingly different models have been proposed in the literature to 
describe the neutral-ionic (N-I) transition first observed in organic mixed-stack
charge-transfer salts.\cite{torr_01,torr_02} Assuming that the coupling between 
the stacks is weak, the system is modelled as a linear chain in which two kinds of
molecules alternate. Assuming furthermore, that it is sufficient to consider a single 
orbital per site, Torrance and Hubbard \cite{hubb-torr} have suggested that the 
different nature of the two molecules can be attributed to different values of the 
on-site energy, while the on-site Coulomb repulsion $U$ can be taken to be 
identical. Taking into account the finite transfer integral between neighboring
molecules we are led to the so-called ionic Hubbard model described 
by the Hamiltonian
\begin{equation}   \begin{split}
      {\mathcal H} &= t \sum_{i\sigma} \left ( c^\dagger_{i\sigma} 
    c^{\phantom\dagger}_{i+1\sigma} + c^\dagger_{i+1\sigma} 
     c^{\phantom\dagger}_{i\sigma} \right )   \\ 
        &  \phantom{=} + U \sum_{i} n_{i\uparrow} n_{i\downarrow} 
       + \frac{\Delta}{2}\sum_{i}(-1)^i n_{i}\,,
\label{eq:ham_ihm}
\end{split}   \end{equation}
where $c^{\dagger}_{i\sigma} (c^{\phantom\dagger}_{i\sigma})$ is the usual 
creation (annihilation) operator of electrons at site $i$ with spin $\sigma$, 
$n_{i\sigma}= c^{\dagger}_{i\sigma}c^{\phantom\dagger}_{i\sigma}$
and $n_i=\sum_\sigma n_{i\sigma}$ is the occupation number at site $i$. 

\begin{figure}[htb]
\unitlength 0.8mm
\begin{tabular}{ccc}
$U<\Delta$ & \rule{6mm}{0mm} & $U>\Delta$ 
\\
\begin{picture}(43,57)(0,0)
\put( 0,42){\line(1,0){8}}
\put( 0,21){\line(1,0){8}}
\put( 0,13){\line(1,0){8}}
\put(21,13){\line(1,0){8}}
\put(21,10){\line(1,0){8}}
\put(21, 5){\line(1,0){8}}
\put( 4,42){\makebox(0,0){\footnotesize $\uparrow\downarrow$}}
\put( 4,21){\makebox(0,0){\footnotesize $\uparrow (\downarrow)$}}
\put(25,10){\makebox(0,0){\footnotesize $\uparrow\downarrow$}}
\put(25, 5){\makebox(0,0){\footnotesize $\uparrow (\downarrow)$}}
\put(10,42){\makebox(0,0)[l]{\scriptsize $\Delta+U$}}
\put(10,21){\makebox(0,0)[l]{\scriptsize $\Delta /2$}}
\put(10,13){\makebox(0,0)[l]{\scriptsize $0$}}
\put(31,13){\makebox(0,0)[l]{\scriptsize $0$}}
\put(31,10){\makebox(0,0)[l]{\scriptsize $-\Delta+U$}}
\put(31, 5){\makebox(0,0)[l]{\scriptsize $-\Delta /2$}}
\put( 0, 0){\makebox(0,0)[lt]{\scriptsize Even sites}}
\put(21, 0){\makebox(0,0)[lt]{\scriptsize Odd sites}}
\end{picture}
&&
\begin{picture}(43,57)(0,0)
\put( 0,50){\line(1,0){8}}
\put( 0,21){\line(1,0){8}}
\put( 0,13){\line(1,0){8}}
\put(21,13){\line(1,0){8}}
\put(21,18){\line(1,0){8}}
\put(21, 5){\line(1,0){8}}
\put( 4,50){\makebox(0,0){\footnotesize $\uparrow\downarrow$}}
\put( 4,21){\makebox(0,0){\footnotesize $\uparrow (\downarrow)$}}
\put(25,18){\makebox(0,0){\footnotesize $\uparrow\downarrow$}}
\put(25, 5){\makebox(0,0){\footnotesize $\uparrow (\downarrow)$}}
\put(10,50){\makebox(0,0)[l]{\scriptsize $\Delta+U$}}
\put(10,21){\makebox(0,0)[l]{\scriptsize $\Delta /2$}}
\put(10,13){\makebox(0,0)[l]{\scriptsize $0$}}
\put(31,13){\makebox(0,0)[l]{\scriptsize $0$}}
\put(31,18){\makebox(0,0)[l]{\scriptsize $-\Delta+U$}}
\put(31, 5){\makebox(0,0)[l]{\scriptsize $-\Delta /2$}}
\put( 0, 0){\makebox(0,0)[lt]{\scriptsize Even sites}}
\put(21, 0){\makebox(0,0)[lt]{\scriptsize Odd sites}}
\end{picture}
\end{tabular}
\caption{Energy levels of the ionic Hubbard model in the atomic limit for
$U < \Delta$ and $U > \Delta$.}
\label{fig:s1_ionic_states}
\end{figure}
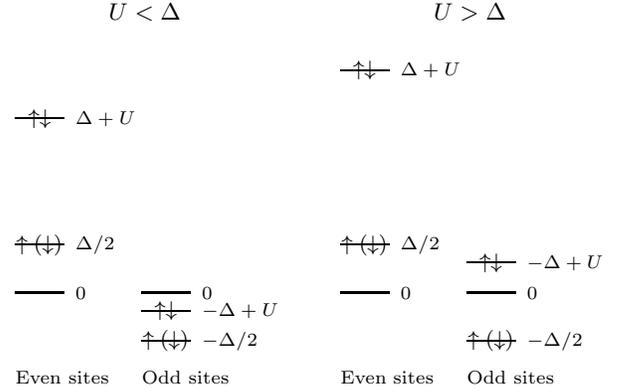

When the number of electrons is exactly equal to the number of sites, the 
competition between the on-site energy difference $\Delta$ and the 
Coulomb energy $U$ will determine whether the system is a band insulator (BI)
or a correlated Mott insulator (MI). It is easily seen by looking at the energy levels in 
the atomic limit, shown in Fig.~\ref{fig:s1_ionic_states}, that if hopping processes
can be neglected, it is energetically favorable to have doubly occupied odd sites 
and empty even sites when $U < \Delta$ (the energy of the pair is 
$E=U-\Delta < 0$), while in the opposite case ($U > \Delta$) in the lowest-energy
configuration every site is occupied by one electron (the energy of such a pair is 
$E=0$), but their spin can be arbitrarily oriented. In the band-insulator state both the 
charge and spin gaps are finite, while in the correlated Mott insulator only the 
charge gap is finite. The transition between these phases takes place at $U = \Delta$. 
 
As was first pointed out by Fabrizio \emph{et al.},\cite{fabri} for finite values 
of the hopping integral $t$ the transition between these two states occurs in two 
steps. The charge gap closes at a critical value $U_{\rm c1}$, but it reopens 
immediately, while the spin gap vanishes at a different value $U_{\rm c2} > 
U_{\rm c1}$. The transition at $U=U_{\rm c1}$ is of Ising-like while the one 
at $U_{\rm c2}$ is a Berezinskii-Kosterlitz-Thouless (BKT) transition. A 
dimerized bond-order (BO) phase exists between the two critical values. 

Since then the model has been studied in detail by several groups using both 
analytic and numerical methods. The most recent works are in 
Refs.~[\onlinecite{torio,zhang,kampf,manma,soos,aligia,leo,legeza}] 
where further references can be found. After some controversy, by now consensus
seems to emerge, the numerical results\cite{manma,leo,legeza} seem to 
support the picture with two transitions. The model has also been extended to 
include intersite Coulomb interaction \cite{bruins,aligia} with rather similar results. 

In realistic charge-transfer salts, however, the intra\-mole\-cular Coulomb energy is 
presumably the largest energy and it is reasonable to assume that its unique role is 
to forbid doubly ionized molecules. If that is so, the transition between the neutral 
and ionic phases is driven by other couplings. In the model studied in detail by 
Avignon \emph{et al.},\cite{avignon} Girlando and Painelli,\cite{girlando} and  
Horovitz and one of the authors \cite{horovitz} it is assumed that donor 
and acceptor molecules alternate along the chain. In its neutral state (D$^0$), the 
highest occupied molecular orbital (HOMO) of the donor is filled by two electrons 
of opposite spin. The ionization energy to remove an electron from this orbital and 
to create thereby a singly ionized D$^+$ donor is $I$. Taking into account the 
Coulomb repulsion $U_{\rm D}$ the energy of the doubly ionized (D$^{2+}$) state 
is $2I + U_{\rm D}$. The donor molecules can thus be described by the Hamiltonian
\begin{equation}    \begin{split}
\label{eq:hd}
{\mathcal H}_{\rm D}  & =  I    \rule{-2mm}{0mm}\sum_{\rm donor\, sites}
       \rule{-2mm}{0mm} (2-n_i)  \\ 
       &  \phantom{=,}+ U_{\rm D} \rule{-2mm}{0mm} \sum_{\rm donor\, sites}
      \rule{-2mm}{0mm} (1-n_{i\uparrow})(1-n_{i\downarrow})\,.
\end{split}   \end{equation}
The energies are measured with respect to the neutral configuration. If 
$U_{\rm D} \gg I$, its role is to forbid doubly ionized sites, empty donor orbitals. 

\begin{figure}[htb]
\unitlength 0.8mm
\begin{picture}(60,35)(0,0)
\put( 0,32){\line(1,0){8}}
\put( 0,15){\line(1,0){8}}
\put( 0, 9){\line(1,0){8}}
\put(33, 9){\line(1,0){8}}
\put(33,27){\line(1,0){8}}
\put(33, 5){\line(1,0){8}}
\put( 4, 9){\makebox(0,0){\footnotesize $\uparrow\downarrow$}}
\put( 4,15){\makebox(0,0){\footnotesize $\uparrow (\downarrow)$}}
\put(37,27){\makebox(0,0){\footnotesize $\uparrow\downarrow$}}
\put(37, 5){\makebox(0,0){\footnotesize $\uparrow (\downarrow)$}}
\put(10,32){\makebox(0,0)[l]{\scriptsize $2I+U_{\rm D}$}}
\put(10,15){\makebox(0,0)[l]{\scriptsize $I$}}
\put(10, 9){\makebox(0,0)[l]{\scriptsize $0$}}
\put(43, 9){\makebox(0,0)[l]{\scriptsize $0$}}
\put(43,27){\makebox(0,0)[l]{\scriptsize $-2A+U_{\rm A}$}}
\put(43, 5){\makebox(0,0)[l]{\scriptsize $-A$}}
\put( 0, 0){\makebox(0,0)[lt]{\scriptsize Donor}}
\put(33, 0){\makebox(0,0)[lt]{\scriptsize Acceptor}}
\end{picture}
\caption{Atomic energy levels of donor and acceptor molecules.}
\label{fig:s1_ionic_levels}
\end{figure}
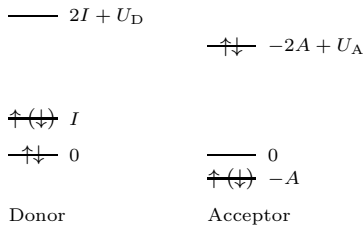

On the other hand the lowest unoccupied molecular orbital (LUMO) of the 
acceptor is empty in its neutral state (A$^0$). The energy is lowered to $-A$, 
where $A$ is the affinity, when the level becomes singly occupied, singly ionized 
(A$^-$), while the doubly ionized (A$^{2-}$) acceptors have a high energy due 
to the Coulomb repulsion $U_{\rm A}$. The acceptor molecules are described 
by the Hamiltonian
\begin{equation}    
 \label{eq:ha}
{\mathcal H}_{\rm A}  =  -A  \rule{-2mm}{0mm}\sum_{\rm acceptor\, sites}
    \rule{-2mm}{0mm} n_i +  U_{\rm A} \rule{-2mm}{0mm}
      \sum_{\rm acceptor\, sites} \rule{-2mm}{0mm} 
      n_{i\uparrow} n_{i\downarrow}\,.
\end{equation}
If $U_{\rm A} \gg A$, its role is to forbid doubly ionized sites, doubly occupied
acceptor orbitals.  The atomic energy levels of the donor and acceptor molecules are 
shown schematically in Fig.~\ref{fig:s1_ionic_levels}. 

Since $I$ and $A$ are fixed internal parameters of the molecules, and it is assumed 
that $I-A > 0$, the molecules are neutral in the ground state of the stack unless 
a transition to an ionized state is driven by the intermolecular Coulomb attraction 
$V$ between the oppositely ionized donors (D$^{+}$) and acceptors (A$^{-}$),
\begin{equation}   \begin{split}
\label{eq:h_c}
 {\mathcal H}_{\rm C}  &=  -V \rule{-2mm}{0mm}\sum_{\rm donor\, sites}
      \rule{-2mm}{0mm} (2-n_i) n_{i+1} \\
       & \phantom{=,} -V \rule{-2mm}{0mm}\sum_{\rm acceptor\, sites} 
        \rule{-2mm}{0mm} n_i(2-n_{i+1}) \,.
\end{split}    \end{equation}     
In this atomic limit this transition should occur at $V = I-A$.

However, in addition to these terms one has to take into account the charge 
transfer between the donor and acceptor molecules described by
\begin{equation}
\label{eq:h_ct}
  {\mathcal H}_{\rm CT} =  t \rule{-2mm}{0mm}\sum_{{\rm all\, sites,\,}\sigma}
       \rule{-2mm}{0mm}  \left( c^{\dagger}_{i \sigma} 
        c^{\phantom\dagger}_{i+1 \sigma}
        +  c^{\dagger}_{i+1 \sigma} c^{\phantom\dagger}_{i \sigma} \right) \,.
\end{equation}
Since for finite hopping the molecules are always at least partially ionized, this may 
smear out the sharp transition between the neutral and ionic phases.

In fact, exact diagonalization on relatively short chains \cite{horovitz} and 
valence-bond calculations \cite{soos-maz,girlando} indicated that the transition 
remains of first order until $t \leq 0.15 (I-A)$. For larger $t$ values a 
second-order or BKT-like transition was observed. The charge, 
spin and charge-transfer gaps seemed to vanish at the same critical intermolecular 
Coulomb coupling $V_{\rm c}$, but the charge gap reopens again.

Not only the interaction driving the neutral-ionic transition is different in the
two models but also the character of the transition seems to be different. The aim 
of the present work is to perform a more careful study of the neutral-ionic transition. 
First it is shown that a unified model can be derived which is identical to the 
donor-acceptor model in the limit when the intramolecular Coulomb repulsion
forbids doubly ionized donors and acceptors, and at the same time it is also
a good approximation to the ionic Hubbard model when $t \ll \Delta$, but $U-\Delta$
can be arbitrary. Detailed numerical calculations are performed on this unified model 
with the help of the density-matrix renormalization-group (DMRG) \cite{white} 
procedure calculating excitation gaps, ionicity, single-site entropy, two-site entropy,  
and the dimer order parameter on long chains and the unified phase diagram is 
determined.

The setup of the paper is as follows. In Sec.~II we transform the Hamiltonian of the 
two models into an effective spin-1 model and show their relationship. In Sec.~III 
we discuss how the quantum phase transition between the neutral and ionic
phases can be best described and our numerical procedure is presented. The 
results are given in Sec.~IV, and the conclusions are drawn in Sec.~V.

\section{Unified spin-1 Hamiltonian}

As mentioned earlier, in the donor-acceptor model, the on-site Coulomb couplings 
are large compared to the other characteristic energies in the problem, and the 
limit $U_{\rm D},\, U_{\rm A} \rightarrow \infty$ can be taken whereby the doubly 
ionized molecular states (empty donor D$^{2+}$ and doubly occupied acceptor 
A$^{2-}$) are forbidden. Since only three states per site survive, this model can be 
mapped onto an effective $S=1$ spin model.\cite{horovitz} The appropriate mapping 
between the allowed electronic states and spin  states is shown in 
Fig.~\ref{fig:mapping_da}. 

\begin{figure}[htb]
\begin{tabular}{ccc}
\begin{tabular}{rcl}
$\left|\, 0                \,\right>$ & $\longrightarrow$ & excluded                \\[1mm]
$\left|\, \uparrow         \,\right>$ & $\longrightarrow$ & $\left|\,  1 \,\right>$ \\[1mm]
$\left|\, \downarrow     \,\right>$ & $\longrightarrow$ & $\left|\, -1 \,\right>$ \\[1mm]
$\left|\, \uparrow\downarrow \,\right>$ & $\longrightarrow$ & $\left|\,  0 \,\right>$ 
\end{tabular}
&
\rule{5mm}{0mm}
&
\begin{tabular}{rcl}
$\left|\, 0                  \,\right>$ & $\longrightarrow$ & $\left|\,  0 \,\right>$ \\[1mm]
$\left|\, \uparrow         \,\right>$ & $\longrightarrow$ & $\left|\,  1 \,\right>$ \\[1mm]
$\left|\, \downarrow     \,\right>$ & $\longrightarrow$ & $\left|\, -1 \,\right>$ \\[1mm]
$\left|\, \uparrow\downarrow \,\right>$ & $\longrightarrow$ & excluded
\end{tabular}
\\
& &
\\[-1mm]
\textrm{Donor}
& &
\textrm{Acceptor}
\end{tabular}
\caption{Mapping of the allowed electronic states of donor and acceptor 
molecules to $S=1$ spin states.} 
\label{fig:mapping_da}
\end{figure}

As is seen, the transfer of an electron with spin $\uparrow$ or $\downarrow$ from 
the neutral donor at site $2j-1$ to the empty, neutral acceptor at site $2j$, the 
process D$_{2j-1}^0$A$_{2j}^0 \rightarrow $ D$_{2j-1}^+$A$_{2j}^-$ 
corresponds in the spin language to an exchange process $\left|0\right>_{2j-1}
 \left|0\right>_{2j} \rightarrow \left|-1\right>_{2j-1} \left|1\right>_{2j}$ or 
$\left|0\right>_{2j-1} \left|0\right>_{2j} \rightarrow \left|1\right>_{2j-1} 
\left|-1\right>_{2j}$. The opposite processes are also possible. No hopping 
could, however, take place between a neutral donor and an ionized acceptor 
or an ionized donor and a neutral acceptor or between ionized neighbors
when the spins of the electrons are parallel. 

However, one has to take into account that due to the fermionic nature 
of the electrons, these hopping processes appear with different signs. 
Acting by the charge-transfer term on a neutral donor-acceptor pair
\begin{equation}    \begin{split}
      {\mathcal H}_{\rm CT} &\, c_{2j-1 \uparrow}^{\dagger} 
       c_{2j-1 \downarrow}^{\dagger} | 0 \rangle_{2j-1} |0\rangle_{2j} \\
     = & \, t \big( c_{2j-1 \uparrow}^{\dagger} c_{2j \downarrow}^{\dagger} -
     c_{2j-1 \downarrow}^{\dagger} c_{2j \uparrow}^{\dagger} \big)
         | 0 \rangle_{2j-1} |0\rangle_{2j} \\
     = & \, t \big(  |\!\uparrow \rangle_{2j-1} |\!\downarrow \rangle_{2j} -
        | \!\downarrow \rangle_{2j-1} | \!\uparrow \rangle_{2j} \big) \,.
\end{split}    \end{equation} 
Similarly when the charge-transfer term acts on an A$_{2j}^0$D$_{2j+1}^0$
pair
\begin{equation}    \begin{split}
      {\mathcal H}_{\rm CT} & \, | 0 \rangle_{2j}  c_{2j+1 \uparrow}^{\dagger} 
       c_{2j+1 \downarrow}^{\dagger} | 0 \rangle_{2j+1} \\
     = & \, t \big( c_{2j \uparrow}^{\dagger} c_{2j+1 \downarrow}^{\dagger} -
     c_{2j \downarrow}^{\dagger} c_{2j+1 \uparrow}^{\dagger} \big)
         | 0 \rangle_{2j} |0\rangle_{2j+1} \\
    = & \, t \big(  | \!\uparrow \rangle_{2j} | \!\downarrow \rangle_{2j+1} -
        | \!\downarrow \rangle_{2j} | \!\uparrow \rangle_{2j+1} \big) \,.
\end{split}    \end{equation} 

The charge-transfer processes can thus be described in the spin language 
by the Hamiltonian
\begin{equation}     \begin{split}
\label{eq:ham_ct_eff}
   {\cal H}_{\rm CT}& = - {\textstyle\frac{1}{2}}t \rule{-2mm}{0mm}
       \sum_{\rm all\, sites}\rule{-2mm}{0mm} S^z_i S^z_{i+1} 
      \left( S^+_i S^-_{i+1} -  S^-_i S^+_{i+1}  \right)   \\
     &  \phantom{=,}  + {\textstyle\frac{1}{2}}t \rule{-2mm}{0mm}
     \sum_{\rm all\, sites} \rule{-2mm}{0mm}   \left( S^+_i S^-_{i+1} 
      - S^-_i S^+_{i+1} \right) S^z_i S^z_{i+1}\,,
\end{split}  \end{equation}
where $S^+_i$ and $S^-_i$ are the usual raising and lowering operators 
corresponding in the fermionic picture to removing or adding an electron. $S^z_i$ 
measures the state of the ion, and the product $S^z_i S^z_{i+1}$ makes sure that 
no charge transfer takes place between a neutral and an ionized site.

Using the commutation relations of the spin operators, this part of the 
Hamiltonian can be written in the form
\begin{equation}   \begin{split}
  {\mathcal H}_{\rm CT}  &= - {\textstyle\frac{1}{2}}t \sum_i S_i^z 
       \left( S_i^+ S_{i+1} ^- + S_i^- S_{i+1}^+ \right)    \\
& \phantom{= ,} - {\textstyle\frac{1}{2}} t  
     \sum_i \left( S_i^+ S_{i+1} ^- + S_i^- S_{i+1}^+ \right) S_{i+1}^z  \,.
\end{split}    \end{equation}

It is easy to check that the same matrix elements are obtained if the other
terms of the Hamiltonian are written in the spin language as
\begin{equation}     \begin{split}
    {\cal H}_{\rm D}  & = I  \rule{-2mm}{0mm}\sum_{\rm donor\, sites}
      \rule{-2mm}{0mm} (S^z_i)^2  \,, \\
    {\cal H}_{\rm A}  &=  -A \rule{-2mm}{0mm}\sum_{\rm acceptor\, sites} 
     \rule{-2mm}{0mm} (S^z_i)^2\,,\\
   {\cal H}_{\rm C}   & = -V \rule{-2mm}{0mm}\sum_{\rm all\, sites} 
     \rule{-2mm}{0mm} (S^z_i)^2 (S^z_{i+1})^2\,.
\end{split}   \end{equation}
Due to charge conservation 
\begin{equation}
     \sum_{\rm donor\, sites} (S^z_i)^2=\sum_{\rm acceptor\, sites} (S^z_i)^2\,,
\end{equation}
and therefore 
\begin{equation}
    {\cal H}_{\rm D} + {\mathcal H}_{\rm A} = {\textstyle\frac{1}{2}}(I-A)
      \sum_{\rm all\,sites}(S^z_i)^2\,.
\label{eq:H_DA}
\end{equation}

We mention that the effective spin-1 Hamiltonian used in Ref.~[\onlinecite{horovitz}]
differs somewhat from the one given above. The charge-transfer term was 
written in the form 
\begin{equation}     \begin{split}
\label{eq:ham_ct_eff-1}
   {\cal H}_{\rm CT}& = -{\textstyle\frac{1}{2}} t \rule{-2mm}{0mm}
      \sum_{\rm all\, sites} \rule{-2mm}{0mm} S^z_i S^z_{i+1} 
         \left( S^-_i S^+_{i+1} + S^+_i S^-_{i+1}  \right)   \\
     &  \phantom{=,}  -{\textstyle \frac{1}{2}}t \rule{-2mm}{0mm}
      \sum_{\rm all\, sites} \rule{-2mm}{0mm}  
       \left( S^-_i S^+_{i+1} + S^+_i S^-_{i+1} \right) S^z_i S^z_{i+1}\,.
\end{split}  \end{equation}
The two expressions can be related by a unitary transformation, by introducing 
phase factors in the mapping, which alternate with a four-site periodicity. The 
appropriate mapping is given in Fig.~\ref{fig:map-phase}. 

\begin{figure}[htb]
\begin{tabular}{ccc}
\begin{tabular}{rcl}
$\left|\, 0                  \,\right>$ & $\longrightarrow$ & excluded                \\[1mm]
$\left|\, \uparrow           \,\right>$ & $\longrightarrow$ & $\left|\,  1 \,\right>$ \\[1mm]
$\left|\, \downarrow         \,\right>$ & $\longrightarrow$ & $\left|\, -1 \,\right>$ \\[1mm]
$\left|\, \uparrow\downarrow \,\right>$ & $\longrightarrow$ & $\left|\,  0 \,\right>$ 
\end{tabular}
&
\rule{9mm}{0mm}
&
\begin{tabular}{rcl}
$\left|\, 0         \,\right>$ & $\longrightarrow$ & $\phantom{-}\left|\,  0 \,\right>$ \\[1mm]
$\left|\, \uparrow           \,\right>$ & $\longrightarrow$ & $-\left|\,  1 \,\right>$ \\[1mm]
$\left|\, \downarrow      \,\right>$ & $\longrightarrow$ & $\phantom{-}\left|\, -1 \,\right>$ 
       \\[1mm]
$\left|\, \uparrow\downarrow \,\right>$ & $\longrightarrow$ & excluded
\end{tabular}
\\
& &
\\[-1mm]
\textrm{Donor at site $4j-3$}
& &
\textrm{Acceptor at site $4j-2$}
\\
& &
\\[1mm]
\begin{tabular}{rcl}
$\left|\, 0                  \,\right>$ & $\longrightarrow$ & excluded                \\[1mm]
$\left|\, \uparrow           \,\right>$ & $\longrightarrow$ & $-\left|\,  1 \,\right>$ \\[1mm]
$\left|\, \downarrow         \,\right>$ & $\longrightarrow$ & $-\left|\, -1 \,\right>$ \\[1mm]
$\left|\, \uparrow\downarrow \,\right>$ & $\longrightarrow$ & $\phantom{-}\left|\, 
       0 \,\right>$ 
\end{tabular}
&
\rule{9mm}{0mm}
&
\begin{tabular}{rcl}
$\left|\, 0       \,\right>$ & $\longrightarrow$ & $\phantom{-}\left|\,  0 \,\right>$ \\[1mm]
$\left|\, \uparrow       \,\right>$ & $\longrightarrow$ & $\phantom{-}\left|\,  1 \,\right>$ 
    \\[1mm]
$\left|\, \downarrow         \,\right>$ & $\longrightarrow$ & $-\left|\, -1 \,\right>$ \\[1mm]
$\left|\, \uparrow\downarrow \,\right>$ & $\longrightarrow$ & excluded
\end{tabular}
\\
& &
\\[-1mm]
\textrm{Donor at site $4j-1$}
& &
\textrm{Acceptor at site $4j$}
\end{tabular}
\caption{Phase factors in the mapping of the donor-acceptor model to a 
spin-1 model.}
\label{fig:map-phase}
\end{figure}

In their work on the extended ionic Hubbard model Aligia and Batista \cite{aligia} 
have shown that a similar mapping to an effective spin-1 model can be used
in this model as well. When the number of electrons is exactly equal to the number 
of sites, the occupancy of states on even and odd sites is strongly correlated.
Due to charge conservation an empty odd site, which itself has low energy,
implies that an even site is doubly occupied, and the energy of this pair
$E= 2\Delta + U$ is larger than that of a pair of singly occupied even and odd sites 
($E=0$), or when the even site is empty and the odd site is doubly occupied 
($E= U-\Delta$). Near the expected transition, $U \approx \Delta$, doubly 
occupied even sites and empty odd sites can, therefore, be neglected.

With this provisio the electronic states of the ionic Hubbard model can be
mapped to the three $S=1$ states. The convention used is shown in 
Fig.~\ref{fig:mapping_iH}.

\begin{figure}[htb]
\begin{tabular}{ccc}
\begin{tabular}{rcl}
$\left|\, 0                  \,\right>$ & $\longrightarrow$ & $\left|\,  0 \,\right>$ \\[1mm]
$\left|\, \uparrow         \,\right>$ & $\longrightarrow$ & $\left|\,  1 \,\right>$ \\[1mm]
$\left|\, \downarrow    \,\right>$ & $\longrightarrow$ & $\left|\, -1 \,\right>$ \\[1mm]
$\left|\, \uparrow\downarrow \,\right>$ & $\longrightarrow$ & excluded
\end{tabular}
&
\rule{5mm}{0mm}
&
\begin{tabular}{rcl}
$\left|\, 0                  \,\right>$ & $\longrightarrow$ & excluded                \\[1mm]
$\left|\, \uparrow         \,\right>$ & $\longrightarrow$ & $\left|\,  1 \,\right>$ \\[1mm]
$\left|\, \downarrow    \,\right>$ & $\longrightarrow$ & $\left|\, -1 \,\right>$ \\[1mm]
$\left|\, \uparrow\downarrow \,\right>$ & $\longrightarrow$ & $\left|\,  
      0 \,\right>$ \\[1mm]
\end{tabular}
\\
& &
\\[-1mm]
\textrm{Even sites}
& &
\textrm{Odd sites}
\end{tabular}
\caption{Mapping of the electronic states of the ionic Hubbard model to $S=1$ 
spin states for the even and odd sites.} 
\label{fig:mapping_iH}
\end{figure}

The requirement that all matrix elements should be identical in the two representations
leads to the following forms for the three terms of the Hamiltonian of the ionic 
Hubbard model:  
\begin{eqnarray}
\nonumber
{\cal H}_t  &=& -{\textstyle\frac{1}{2}}t \rule{-2mm}{0mm}\sum_{\rm all\, sites}
      \rule{-2mm}{0mm} S^z_i S^z_{i+1} \left( S^+_i S^-_{i+1} - 
         S^-_i S^+_{i+1}  \right)\\
     &&  +{\textstyle\frac{1}{2}}t \rule{-2mm}{0mm}\sum_{\rm all\, sites}
    \rule{-2mm}{0mm} \left( S^+_i S^-_{i+1} - S^-_i S^+_{i+1} \right) 
      S^z_i S^z_{i+1}\,, \\
  {\cal H}_U  &=& U \sum_{\rm odd\, sites}\left( 1-(S_i^z)^2 \right) 
\label{eq:H_U},\\
{\cal H}_\Delta &=& {\textstyle\frac{1}{2}}\Delta \left[\sum_{\rm even\, sites} 
      (S_i^z)^2   -  \sum_{\rm odd\,  sites}  \left( 2  - (S_i^z)^2  \right) \right] \!\!.
\end{eqnarray}

The hopping term has exactly the same form as the charge-transfer term in the
donor-acceptor model. The resemblance of the atomic term of the ionic 
Hubbard model to that of the donor-acceptor model becomes manifest when 
charge conservation is taken into account. The sum of the on-site Coulomb 
term (${\cal H}_U$) and the ionic term (${\cal H}_\Delta$) can be written as
\begin{equation}
{\cal H}_U + {\cal H}_\Delta = -\frac{\varepsilon}{2}\sum_i (S_i^z)^2 + 
\frac{N\varepsilon}{2}
\end{equation}
where $N$ is the number of sites and the parameter
\begin{equation}
     \varepsilon=U-\Delta 
\label{eq:eps_ihm}
\end{equation}
has been introduced. Apart from a constant term this is the same as the atomic 
part of the donor-acceptor model given in Eq. (\ref{eq:H_DA}), if the identification
\begin{equation}
       \varepsilon = - (I - A)
\label{eq:eps_da}
\end{equation}
is done. Thus in this representation the Hamiltonian of the donor-acceptor model 
is just an extension of that of the ionic Hubbard model, including the nearest-neighbor 
Coulomb interaction. Even sites in the ionic Hubbard model correspond to acceptors
and odd sites to donors.

There is, however, an essential physical difference. While in the ionic Hubbard model
the two transitions occur at positive values of $\varepsilon$, in the donor-acceptor 
model $\varepsilon$ is assumed to be negative, and the nearest-neighbor Coulomb 
coupling is needed to drive the transition. The question we want to address in the 
remaining part of this paper is how the transitions obtained in the ionic Hubbard 
model can be related in this unified model to the first or second order transitions 
of the donor-acceptor model.

\section{Detecting and locating phase transitions}

A customary numerical procedure to detect and locate quantum phase transitions
is to calculate energy gaps or correlation functions. A drawback
of this procedure is that even if DMRG allows to determine these quantities 
on long chains, quite often it is very difficult to draw firm conclusions from 
finite-size scaling. This is well illustrated by the longstanding controversy over 
the existence of two phase transitions in the ionic Hubbard model.

Recently we have proposed\cite{legeza} that the two-site entropy, which is 
easily accessible in DMRG, should be considered when quantum phase transitions
are studied. In this section we present briefly the quantities that 
were used to locate the phase transitions and our method to control the numerical 
accuracy is also summarized.

\subsection{Energy gaps}

The natural quantities to be calculated are the excitation gaps. There are 
several types of excited states which can be considered.

\begin{enumerate}
\item  $\Delta E_{\rm c}$ is the energy needed to add or remove an electron. Since 
originally there is an equal number of $\uparrow$ and $\downarrow$ electrons in the
system, in the spin language, this gap is the energy difference between the lowest 
lying states with total spin $S^z_{\rm T} =\pm 1$ and $0$. Denoting the lowest energy 
in the spin sector $S^z_{\rm T}$ by $E_0(S^z_{\rm T})$, $\Delta E_{\rm c} = 
E_0(S^z_{\rm T}=1) - E_0(S^z_{\rm T}=0)$.

\item  $\Delta E_{\rm s}$ is the energy gap of spin-flip excitations. In the spin language, it 
corresponds to the energy gap between the lowest lying states with total spin $2$ 
and $0$, $\Delta E_{\rm s} = E_0(S^z_{\rm T}=2) - E_0(S^z_{\rm T}=0)$.  

\item  $\Delta E_{\rm CT}$ is the excitation energy needed to transfer charge from a 
donor to an acceptor without changing the total charge and spin. In the spin language, 
it is the energy difference between the two lowest lying states of the 
$S^z_{\rm T}=0$ spin sector, $\Delta E_{\rm CT} = E_1(S^z_{\rm T}=0) - 
E_0(S^z_{\rm T}=0)$. When calculating this quantity one has
to take into account that for an open chain with an even number of lattice sites 
the two lowest lying singlet states are separated by a finite energy gap even in 
the thermodynamic limit, if the system is dimerized. In order to get the proper 
charge-transfer gap the finite-chain calculations have to be done with periodic
boundary condition.  
\end{enumerate}

In order to facilitate comparison with the work by Manmana 
\emph{et al.}\cite{manma} where the behavior of the gaps of the nontruncated
ionic Hubbard model have been carefully studied, we note that their $\Delta_1$
is twice our charge gap, $\Delta_1 \equiv 2 \Delta E_{\rm c}$, the spin
gap is definied in the same way, $ \Delta_{\rm S} \equiv \Delta E_{\rm s}$,
while the quantity we characterized as charge-transfer gap is called by them
exciton gap, $\Delta_{\rm E} \equiv \Delta E_{\rm CT}$.

\subsection{Ionicity and dimer order}

Alternatively one can look at an appropriately defined order parameter. One
such quantity could be the average charge on the molecules, the ionicity. In the 
spin representation it is given by the expectation value of $(S^z_i)^2$, 
\begin{equation}
        \varrho_i \equiv \langle \Psi_{\rm GS}|(S^z_i)^2| \Psi_{\rm GS}\rangle\,,
\end{equation}
where $| \Psi_{\rm GS} \rangle$ is the ground state wave function. It turns out 
that this is a good quantity to look at when the transition is of first order and 
the ionicity has a finite jump at the transition point. In a second-order
transition, however, the ionicity is continuous, although with an infinite derivative
in the thermodynamic limit, so it is difficult to get reliable information from its 
behavior in finite systems.

As has been pointed out in Refs.~[\onlinecite{horovitz,manma}], another proper 
order parameter could be the dimer order parameter $D$. An indication of the 
existence of dimer order can be obtained by measuring the alternation in the 
bond energy 
\begin{equation}
     D_i = \langle{\cal H}_{i-1,i}\rangle - \langle{\cal H}_{i,i+1}\rangle
\end{equation}
in the middle of a long enough open chain. The dimer order parameter can thus 
be defined as 
\begin{equation}
D = \lim_{N\to\infty} |D_{N/2}| \,.
\label{eq:dimerord}
\end{equation}

\subsection{Single-site and two-site entropies}

Knowing the  reduced density matrix $\rho_i$ of site $i$, which is obtained
from the wave function of the total system by tracing out all configurations of
all other sites, the von Neumann entropy of this site can be determined from
$s_i = -{\rm Tr} \rho_i \ln \rho_i$. For a model with $q$ degrees of freedom
per site $s_i$ may vary between 0 and $\ln q$. In a chain with free ends, 
the site entropy at the center of the chain $s_{N/2}$ is free from end effects
if the chain length is larger than the coherence length, and the anomalies
appearing in this quantity as a function of the coupling constant can be used 
to detect quantum phase transitions. The site entropy has a jump at a 
first-order transition, while if a second-order transition takes place between 
two differently ordered phases, the site entropy is expected to show a sharp 
maximum at the transition point. A similar quantity, the concurrence, can also 
be used to locate quantum phase transition points, as has been shown by 
Vidal {\sl et al.}\cite{vidal} for the Ising model in transverse field. 

The single-site entropy, however, is not always a good indicator of a quantum phase
transition. It can be featureless when it is insensitive to the breaking of symmetry 
that distinguishes the two phases. Recently it has been pointed out\cite{legeza} that 
in such cases quantum phase transitions can be more efficiently detected and 
located by studying the von Neumann entropy of an ensemble of two lattice sites 
in the middle of a long chain. The quantity defined by
\begin{equation}
     s_{i,i+1} = -{\rm Tr} \rho_{i,i+1} \ln \rho_{i,i+1}\,,
\end{equation}
where $\rho_{i,i+1}$ is the reduced density matrix of two neighboring sites,
may display---when taken in the middle of the chain, at $i = N/2$---a relatively 
sharp maximum at a phase transition point even though the single-site entropy does 
not.

When the system gets dimerized at the transition, the breakdown of translational 
symmetry can also be detected by calculating---as an alternative to the usual 
dimer order parameter---the difference of two-site entropies on neighboring sites,
in the center of the chain
\begin{equation}
     D_s = s_{i+1,i+2}-s_{i,i+1}  \qquad   i=N/2 \,.
\end{equation}

\subsection{Numerical accuracy}

The numerical calculations were performed on finite spin chains up to $N=800$ lattice 
sites when open boundary condition (OBC) was used and up to $N=256$ sites for 
periodic boundary conditions (PBC) using the DMRG technique,\cite{white} and 
the dynamic block-state selection (DBSS)  approach.\cite{legeza02,legeza03} 
All data shown in the figures were obtained with OBC unless stated otherwise.
We have set the threshold value of the quantum information loss $\chi$ to $10^{-8}$ 
and the minimum number of block states $M_{\rm min}$ to $256$. The
maximum number of block states varied in the range $300-800$ for OBC and
$500-2000$ for PBC, respectively. All eigenstates of the model have been targeted 
independently using two or three DMRG sweeps. The entropy sum rule was 
checked for all finite chain lengths for each DMRG sweep, and it was found that 
the sum rule was satisfied after the second sweep already. 

After accomplishing the infinite lattice procedure and using White's wave-function
transformation method \cite{stvec} the largest value of the fidelity error of the 
starting vector of the superblock diagonalization procedure, 
$\delta\varepsilon_{\Psi_{\rm stv}} = 1- \langle \Psi_{\rm T}|\Psi_{\rm stv} \rangle$, 
where $\Psi_{\rm stv}$ is the starting vector and $\Psi_{\rm T}$ is the target state 
determined by the diagonalization of the superblock Hamiltonian, was of the order 
of $10^{-9}$. 

As another test of the accuracy we calculated the ionicity $(\varrho_i)$, lattice site 
entropy ($s_i$), two-site entropy ($s_{i,i+1}$) and dimer order $(D_i)$ profile using 
PBC. In principle, each of them should be independent of $i$ for a given finite chain 
length. Using the DBSS approach with $\chi=10^{-6}$, $M_{\rm min}=256$ for
chains up to $N=256$ sites the values obtained differ from their mean by less than 
$10^{-5}$ for all sites.

It is important to emphasize that when a maximum or cusp is looked for in the 
single-site or two-site entropies, the application of the DBSS approach is 
crucial, otherwise---if the number of block states $m$ were fixed during the 
calculation---a very large number of states would have to be used. This is 
due to the rapid increase of entanglement around the transition points, 
thus a constant $m$ can lead to a significant and non-constant cut in the entropy 
functions.  

In order to obtain the energy gaps, order parameters and entropies in the
$N\rightarrow\infty$ thermodynamic limit, finite-size scaling analysis has been 
performed using scaling functions appropriate for OBC and PBC.\cite{buchta} 

\section{Numerical results}

Since a nontrivial truncation procedure has to be performed to reduce the ionic 
Hubbard model to the unified model with $\varepsilon > 0$ and $V = 0$, first we 
have studied to what extent the truncation modifies the location of the phase 
transition in the model. As a next step we checked the first-order transition
of the donor-acceptor model for $\varepsilon < 0$ and $t \ll V$. Then we 
extended the calculation to intermediate values of the couplings to see how the two
limits can be related. 

\subsection{The ionic Hubbard model versus the unified model at $V=0$}

\begin{figure}[htb]
\includegraphics[scale=0.5]{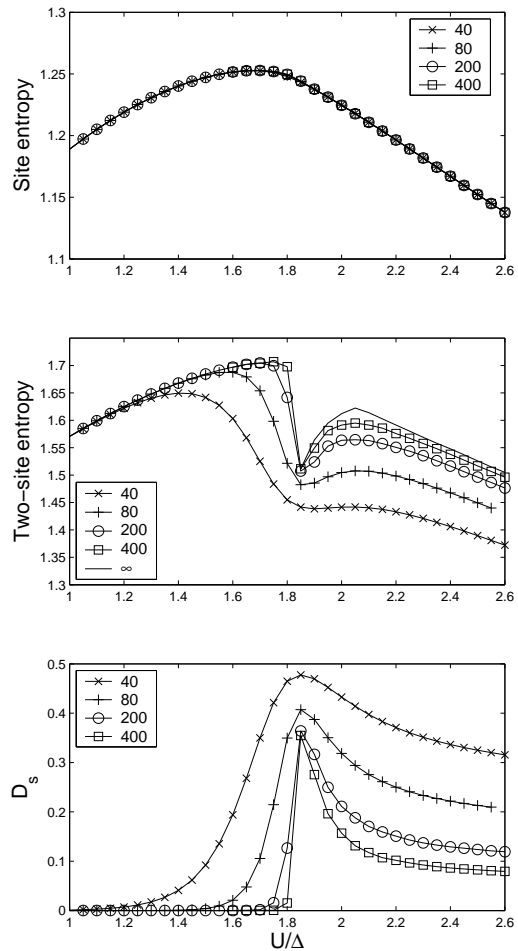}
\caption{Single-site and two-site entropies as well as the dimerization in the
two-site entropy of the ionic Hubbard model for $t/ \Delta= 0.5$ as a function of
$U/\Delta$.}
\label{fig:hubi_t05_entr}
\end{figure}

As mentioned before the ionic Hubbard model given by Eq.~(\ref{eq:ham_ihm}) has
been shown to possess two phase transitions. An ionic bond-ordered dimerized 
phase has been found between the neutral regular and the ionic regular phases. 
Since the location of the BKT transition is an especially difficult problem when
the vanishing of gaps is studied numerically, we have chosen to calculate 
the single-site and two-site entropies as a function of $U$ for various values of $t$ 
by taking $\Delta$ as the unit of energy. As an example we show in 
Fig.~\ref{fig:hubi_t05_entr} our results for various finite chain lengths at 
$t / \Delta = 0.5$.

\begin{figure}
\includegraphics[scale=0.4]{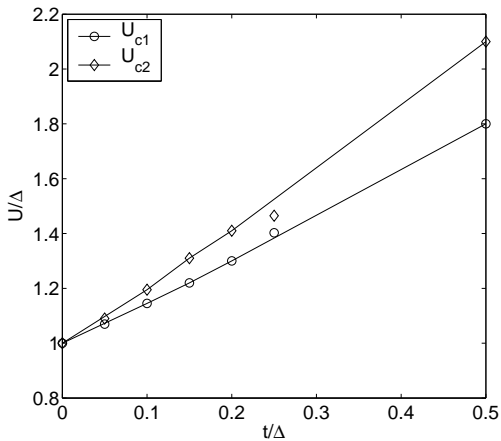}
\caption{Phase diagram of the ionic Hubbard model. The critical values at
$t/\Delta = 0.05$ and $0.25$ are taken from Ref.~[\onlinecite{manma}].}
\label{fig:phase-ionic}
\end{figure}

As seen in the figure the single-site entropy of the central site is a rather smooth 
curve without any cusp or sharp maximum, while the two-site entropy has two 
maxima. As the size of the system is increased the maxima get closer, but 
finite-size scaling analysis shows that they remain separated and develop into 
two cusps in the thermodynamic limit. The same behavior is found for other 
values of $t/\Delta$, except that with decreasing $t$ the two peaks get even closer. 
This result is in perfect agreement with the two-transition scenario mentioned above.
One peak can be identified with $U_{\rm c1}$, the other with $U_{\rm c2}$.
The phase diagram of the ionic Hubbard model obtained from these calculations 
for $t/\Delta \leq 0.5$ is given in Fig.~\ref{fig:phase-ionic}. 

The values obtained by Manmana \emph{et al.}\cite{manma} for $U_{\rm c1}$ 
and the bounds for $U_{\rm c2}$ from their study of the gaps are also shown in the
figure. They fit very nicely on the curves obtained from the peaks of the two-site 
entropy.

A hint about the nature of the phases can be obtained when the alternation of the
two-site entropy $D_s$ is looked at. As seen in the third panel of 
Fig.~\ref{fig:hubi_t05_entr} it vanishes for small and large values of $U/\Delta$
and is finite in the intermediate region only, indicating that in the BO phase, the 
system is spontaneously dimerized, while it is regular in the BI and MI phases. 
The critical value where dimerization first appears coincides well with
$U_{\rm c1}$, while it cannot be established with certainty that the dimer
order disappears exactly at $U_{\rm c2}$, since convergence at the BKT 
transition is very slow. There is no doubt, however, that the dimer order 
disappears at large $U$ values and a consistent picture is obtained if it is
assumed that spontaneous dimerization occurs in a narrow range only, between 
the two transitions.

Furthermore, the phase boundaries allow to determine the range where the 
truncated Hamiltonian can be justified. That is the regime where 
$(U/\Delta)_{\rm c1}$ and $(U/\Delta)_{\rm c2}$ vary linearly with $t/\Delta$, 
i.e., where the relevant parameter is the combination 
$\varepsilon/t \equiv (U-\Delta)/ t $. The slopes of the curves give
\begin{equation}
   (\varepsilon/t)_{\rm c1} \approx 1.4  \,, \qquad (\varepsilon/t)_{\rm c2} \approx  2\,, 
\label{eq:criti-appr}
\end{equation}
and the curvature is rather small up to $t/\Delta \approx 0.5$.

\begin{figure}[htb]
\includegraphics[scale=0.5]{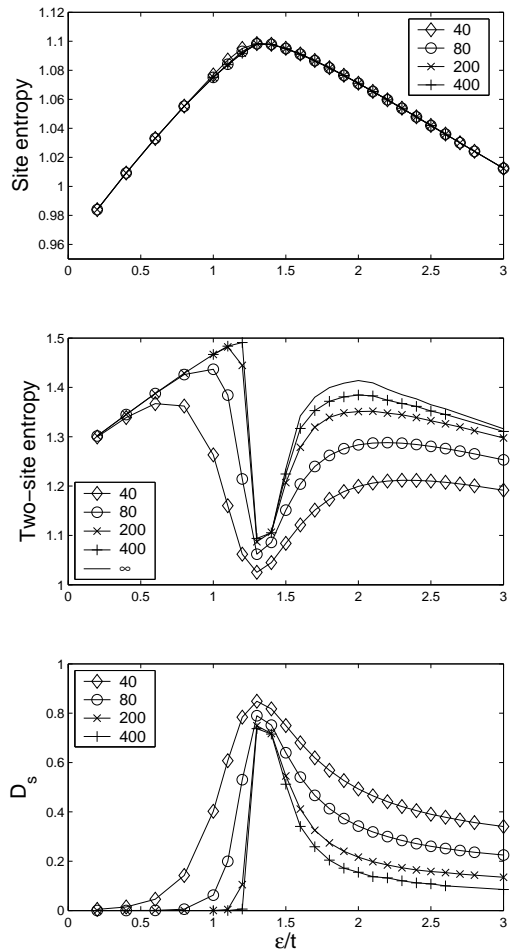}
\caption{Same as Fig.~\ref{fig:hubi_t05_entr} but for the effective spin-1 model 
at $V = 0$ as a function of $\varepsilon/t \equiv (U -\Delta)/t$. } 
\label{fig:v0_entr}
\end{figure}

As a check of these results we have repeated the calculations on the unified model
at $V=0$ for $\varepsilon > 0$. The results for various finite chain lengths are 
shown in Fig.~\ref{fig:v0_entr}.

The site entropy is a rather smooth curve, like for the full model. Even if eventually 
it develops a cusp at $(\varepsilon/t)_{\rm c1}\simeq 1.3$, there is no indication 
in the single-site entropy of the second, BKT-like transition. In contrast to this, 
the two-site entropy exhibits again two maxima. Although for the longest chains 
the entropy curves have been smoothened close to the maxima due to the fact 
that the number of block states selected dynamically has reached the maximum 
number of block states that our code could handle, nevertheless finite-size 
scaling analysis of the curves gave $(\varepsilon/t)_{\rm c1} \approx 1.3$ and 
$(\varepsilon/t)_{\rm c2} \approx 2$ for the two critical values, in agreement 
with our result given in Eq.~(\ref{eq:criti-appr}) and with earlier results on the 
full ionic Hubbard model.\cite{legeza,leo}

Moreover, the two-site entropy $D_s$ shows clearly that the system is
spontaneously dimerized above the first critical value $(\varepsilon/t)_{\rm c1}$, 
but again although $D_s$ certainly vanishes at large enough $\varepsilon/t$, 
the exact location where this happens cannot be determined from the available 
chain lengths using finite-size scaling analysis.

\subsection{Large negative values of $\varepsilon/t$}

The donor-acceptor model corresponds to $\varepsilon < 0$. In the absence of
charge transfer between neighbors $(t=0)$ the transition from the neutral to the 
ionic phase takes place at 
\begin{equation}
    V= {\textstyle\frac{1}{2}}(I - A) \equiv - {\textstyle\frac{1}{2}}\varepsilon \,.
\end{equation}
This is a first-order transition where the ionicity jumps from zero to one and none 
of the gaps go to zero continuously as discussed in Ref.~[\onlinecite{horovitz}]. 
In the neutral state, all sites are in a pure state, the single-site and two-site 
entropies are zero. In the ionic phase the high degeneracy is due to the arbitrary
orientation of the spins, and the site entropy is $\ln 2$, the two-site 
entropy is $\ln 4$. The jump occurs at the transition point. 

\begin{figure}[htb]
\includegraphics[scale=0.5]{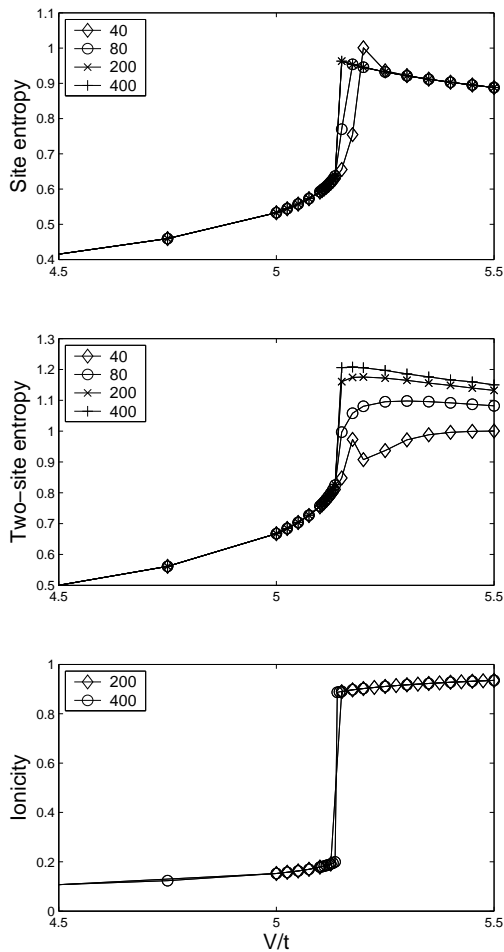}
\caption{The single-site entropy ($s_i$), two-site entropy ($s_{i,i+1}$),
and ionicity ($\varrho_i$) in the middle of the chains for different chain
lengths at $\varepsilon/t= -10$ as a function of $V/t$.}
\label{fig:t_02_entr}
\end{figure}

For finite $t$ the ionicity and the single-site and two-site entropies are nonvanishing
in the neutral phase and their value is less than the maximal one in the ionic phase.
Nevertheless, when $t$ is small enough compared to $V$ and $|\varepsilon|$,
a finite jump can be detected in these quantities near $V \approx -\varepsilon/2$. 
The behavior of these quantities as a function of $V/t$ for
$\varepsilon/t = -10$ is shown in Fig.~\ref{fig:t_02_entr}. 

\begin{figure}[htb]
\includegraphics[scale=0.5]{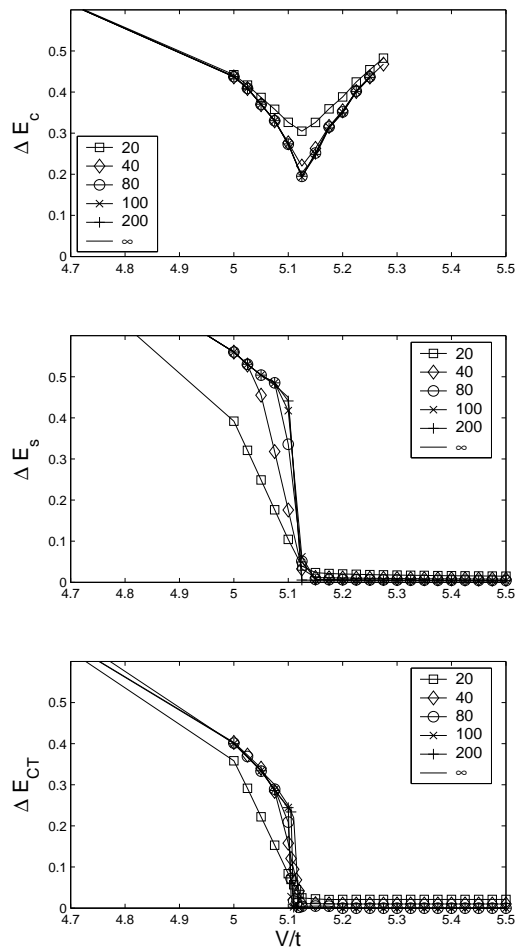}
\caption{Charge gap, spin gap and charge-transfer gap for a small nonvanishing 
charge-transfer integral $(\varepsilon/t=-10)$ using PBC.}
\label{fig:t_02_gaps}
\end{figure}

The excitation energies obtained on finite chains using PBC for $\varepsilon/t=-10$ 
as a function of $V/t$ in the vicinity of the N-I transition are shown in 
Fig.~\ref{fig:t_02_gaps}. Although the charge gap $\Delta E_{\rm c}$ has 
a minimum for all finite $N$, it remains finite even in the thermodynamic limit. 
The minimum in the $N\rightarrow\infty$ limit is at $V/t\simeq 5.13$, the same 
value where the ionicity and the entropies have a jump. At the same point, the 
spin gap $\Delta E_{\rm s}$ and the charge-transfer gap $\Delta E_{\rm CT}$ 
jump to zero and remain zero for larger $V/t$ values. 

The dimer order parameter was found to be of the order of 
$10^{-3}$ or less for all $V/t$ for $N=400$ lattice sites already. Finite-size 
scaling extrapolation gives less than $10^{-4}$ for the dimer order parameter. 

Similar results have been found everywhere for $\varepsilon/t<-8$ except that 
the transition point shifts to larger $V/t$ values with decreasing $\varepsilon/t$ 
as expected. We can, therefore, conclude---in agreement with earlier works 
\cite{girlando,horovitz}---that for small values of $t/V$, or equivalently for
large negative values of $\varepsilon/t$ the system undergoes a first-order transition 
from a regular neutral to a regular ionic phase. 

\subsection{Smaller values of $\varepsilon/t$}

\begin{figure}[htb]
\includegraphics[scale=0.5]{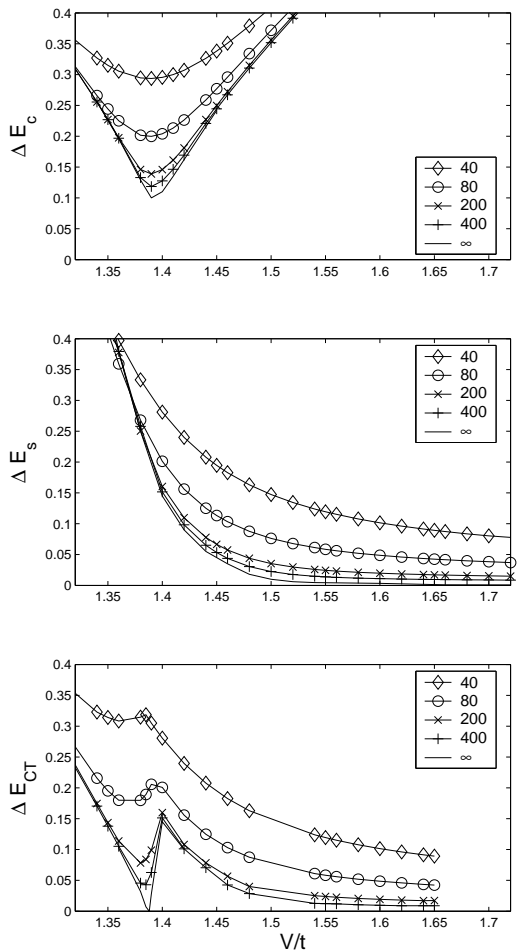}
\caption{Charge gap, spin gap and charge-transfer gap for $\varepsilon/t=-2$.}
\label{fig:t_05_gaps}
\end{figure}

As we have seen, the unified model describes on the one hand the two transitions 
taking place at $V=0$ and $\varepsilon/t > 0$, and on the other hand, the unique 
first-order transition at large negative values of $\varepsilon/t$. The question we 
want to address now is how these two limits can be incorporated into a unified phase 
diagram. We have, therefore, done similar calculations for fixed smaller negative 
and positive values of $\varepsilon/t$ as a function of $V/t$. The excitation 
energies obtained on finite chains using OBC for $\varepsilon/t=-2$ are shown 
in Fig.~\ref{fig:t_05_gaps}.

Comparison to the results obtained in Ref.~[\onlinecite{manma}] shows that
the gaps behave surprisingly similarly to that found in the full ionic Hubbard model. 
The charge-transfer gap vanishes at a critical value $(V/t)_{\rm c1}$ 
which for $\epsilon/t = -2$ is $(V/t)_{\rm c1} \approx 1.39$, 
but reopens again and closes a second time at another critical point, whose 
exact location, however, cannot be determined by finite-size scaling from
the available chain lengths. When the same calculations are performed with 
PBC, the charge-transfer gap vanishes at $(V/t)_{\rm c1}$ and remains zero. 
Thus the reopening of the gap is due to our use of OBC. As we will see, the 
system is spontaneously dimerized for $V/t$ larger than this critical value, and as 
has been discussed in Sec.~III, this gives rise to the finite gap between the two 
lowest lying levels.

The extrapolated spin gap, on the other hand, is finite at $(V/t)_{\rm c1}$ and 
closes at a second critical value $(V/t)_{\rm c2}$. Due to the very
slow decay of the spin gap even on rather long chains,  the exact location of 
the second phase transition could not be determined accurately from the gap. 

The charge gap $\Delta E_{\rm c}$ has a minimum for all $N$ and the 
location of the minimum scales in the thermodynamic limit to 
$(V/t)_{c1}\simeq 1.39$, however, the gap remains finite at this transition 
point. The same behavior has been found for the one-particle gap $\Delta_1$ 
in Ref.~[\onlinecite{manma}].  

The ionicity (not shown) is no longer a good indicator of the neutral-ionic 
transition, it is found to be a continuous function of $V/t$. Instead of that
the dimer order has to be studied, as has been done in the ionic Hubbard 
model. The behavior is again very similar. The extrapolated value of the
measured dimer order parameter is exceedingly small, of the order of 
$10^{-5}-10^{-4}$ for $V/t < (V/t)_{\rm c1}$.
It becomes finite above this transition point, but disappears again at a larger value
of $V/t$. Calculations on very long chains ($N=800$) still do not allow a 
reliable extrapolation to determine the coupling where this happens.

\begin{figure}[htb]
\includegraphics[scale=0.5]{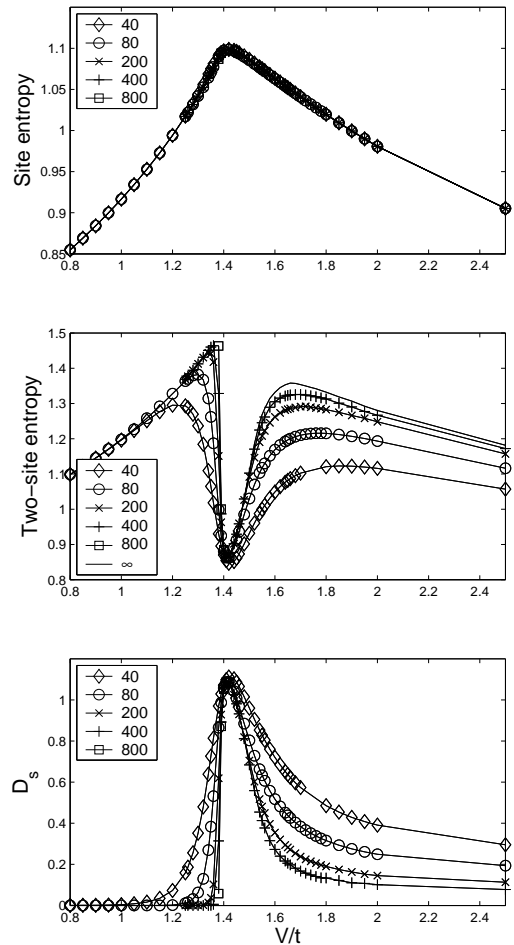}
\caption{The single-site entropy, two-site entropy, and the dimerization of 
the two-site entropy for $\varepsilon/t=-2$.}
\label{fig:t_05_entr}
\end{figure}

Since it is difficult to locate the second transition from the vanishing of the spin gap
or the dimerization, we have studied the behavior of the entropy. Our results obtained 
on finite chains are shown in Fig.~\ref{fig:t_05_entr}.

The single-site entropy seems to develop a cusp indicating a second-order phase 
transition taking place at $(V/t)_{\rm c1}\simeq 1.4$, but only one transition is
found. In contrast to this the two-site entropy possesses two maxima for all finite 
chain lengths and they both develop into a cusp-like peak in the thermodynamic 
limit. The critical points are at $V/t\simeq 1.39$ and $1.66$.  Calculations 
in the vicinity of $V/t=1.39$ on chains up to $N=800$ lattice sites confirm that 
although the rise in the two-site entropy is rather steep, there is no jump in it,
the transition is of second order. 
 
Similar results, two transitions have been found for other values of $\varepsilon/t$ 
as a function of $V/t$, when $\varepsilon/t$ is larger than about $-8$. When
$\varepsilon/t$ is larger than the critical $(\varepsilon/t)_{\rm c1} \approx 1.3$ or
$(\varepsilon/t)_{\rm c2} \approx 2$, one or both transitions appear for negative
$V/t$ values. 

\subsection{Phase diagram}

\begin{figure}[htb]
\includegraphics[scale=0.4]{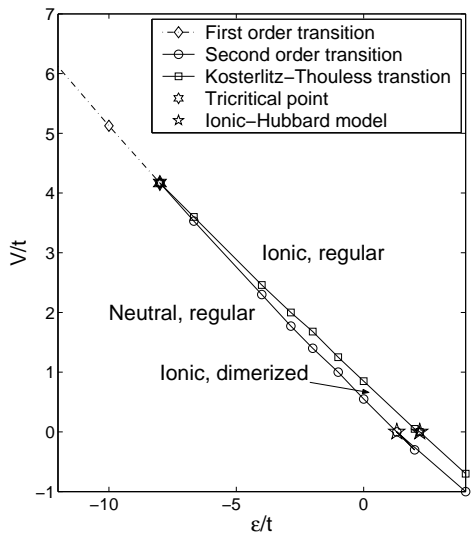}
\caption{Phase diagram of the unified model describing the ionic Hubbard 
and donor-acceptor models as a function of the parameters ${\varepsilon/}t$ 
and $V/t$. Dashed line indicates first-order transition, solid lines correspond to 
second-order or BKT transitions. The star denotes the tricritical point. 
The phase boundaries of the full ionic Hubbard model are also shown.}
\label{fig:phasediagram}
\end{figure}

The results obtained until now can be summarized in the phase diagram shown 
in Fig.~\ref{fig:phasediagram}.  

The two transitions obtained in the ionic Hubbard model and the first-order
transition of the donor-acceptor model appear in fact as two limiting cases of the
unified model. When $\varepsilon$ is negative and $|\varepsilon| \gg t$, the 
neutral-ionic transition is discontinuous, the ionicity jumps by a finite amount.
A tricritical point appears at about $-\varepsilon/t \approx 8$ and 
$V/t \approx 4.2$, beyond which two transitions are found. The charge-transfer 
gap vanishes at one of the transitions, while the spin gap does so at the other one. 

Since the dimer order parameter is nonvanishing in the narrow region only,
between the two continuous transitions, three phases can be identified,
a regular neutral, a regular ionic, and a dimerized phase inbetween.

\section{Conclusion}

In summary, we have studied the neutral-ionic transition in organic mixed-stack 
compounds using the density-matrix renormalization-group method. First, we have 
shown that a unified model can be derived which is identical to the donor-acceptor
model when the intramolecular Coulomb repulsion is the largest energy in the problem
and it represents a good approximation to the ionic Hubbard model when 
$\Delta/t \gg 1$.
 
Detailed numerical calculations have been performed on this unified model calculating 
excitation gaps, ionicity, lattice site entropy and dimer order parameter on long 
chains. We have shown that the best quantity to study is the two-site entropy. It
exhibits the transitions most clearly, even at points where finite-size effects are
important and the vanishing of gaps in the thermodynamic limit is difficult to establish.
In this way we could determine the unified phase diagram as a function of intersite 
Coulomb interaction and the intraatomic energies. The results are in complete 
agreement with earlier results on the ionic Hubbard model, while earlier works on the
donor-acceptor model have found a single (first or second order) transition only.
This is probably due to the closeness of the two transitions, which could not be 
resolved on short chains.   

In the model studied, spontaneous dimerization appears in a narrow range of 
couplings only, while experiments indicate that the ionic phase is dimerized
when electron-phonon coupling is taken into account and the displacement
of ions is permitted. Extension of the calculations in this direction is in progress. 

\acknowledgments

The authors are grateful to R. Noack and L. Tincani for useful discussions.
This research was supported in part by the Hungarian Research Fund(OTKA)
Grants No.\ T 043330 and F 046356. The authors acknowledge
computational support from Dynaflex Ltd under Grant No. IgB-32.
\"O. L. was also supported by the J\'anos Bolyai scholarship.

\end{document}